# Using Full-text Content of Academic Articles to Build a Methodology Taxonomy of Information Science in China

Heng Zhang, Chengzhi Zhang[*]

*Department of Information Management, Nanjing University of Science and Technology, Nanjing 210094, China*

zh_heng@njust.edu.cn，zhangcz@njust.edu.cn

**Abstract**: Research on the construction of traditional information science methodology taxonomy is mostly conducted manually. From the limited corpus, researchers have attempted to summarize some of the research methodology entities into several abstract levels (generally three levels); however, they have been unable to provide a more granular hierarchy. Moreover, updating the methodology taxonomy is traditionally a slow process. In this study, we collected full-text academic papers related to information science. First, we constructed a basic methodology taxonomy with three levels by manual annotation. Then, the word vectors of the research methodology entities were trained using the full-text data. Accordingly, the research methodology entities were clustered and the basic methodology taxonomy was expanded using the clustering results to obtain a methodology taxonomy with more levels. This study provides new concepts for constructing a methodology taxonomy of information science. The proposed methodology taxonomy is semi-automated; it is more detailed than conventional schemes and the speed of taxonomy renewal has been enhanced.

---

# 1.0 Introduction

Collecting the research methodologies of a discipline to construct a methodology taxonomy can broaden researchers' horizons and inspire them to solve a particular problem. Methodology taxonomy can help novice researchers select their research methods, thereby enabling them to find feasible research ideas faster. It is also of great significance to system construction for a special domain (Dick 2019). Forming a methodology taxonomy can also serve as a marker for judging whether a discipline has matured over the years. However, only when a discipline is appropriately developed, can sufficient research methodologies be accumulated to form a methodology taxonomy. Specifically, regarding the information science discipline, there is need for a methodology taxonomy to guide the research on informatics.

Most of the existing research on the construction of a methodology taxonomy of information science is conducted manually. Researchers select representative academic papers on information science. They then summarize the research methodologies from the full-text content (Chu and Ke 2017) or specific parts (such as title, keywords (Lu et al. 2019), and abstracts) of the academic papers via content analysis (Hawkins et al. 2003). Finally, they classify these research methodologies according to a certain strategy to produce the methodology taxonomy. The conventional methods for constructing these taxonomies have some shortcomings in that they are time consuming and laborious, as well as constrained by the number of selected papers (ranging from hundreds to thousands) and limited availability of paper content (such as only titles, keywords, and abstracts). As a result, the research methodologies obtained are not sufficiently comprehensive; the artificial summaries of the research methodologies that are generated are usually abstract and the taxonomies generally only have a three-level structure. There is no doubt that the first level is methodology of information science. The second level varies according to the classification criteria of different researchers, for example, analysis method and data collection method. The third level usually includes abstract conceptual features, such as experimental and statistical methods. Algorithms often used in information science research, such as support vector machines and the naive Bayesian algorithms, are not reflected in these taxonomies. Therefore, it is difficult for researchers to visualize detailed and specific research methodologies using a three-level methodology taxonomy.

With the rise of the open science movement, access to the full-text content of academic papers has become increasingly easy. In this study, we collected such content for information science and constructed a basic three-level methodology taxonomy via manual annotation. The research content and methodology sections of the academic papers contain descriptions of the research methodologies. We used the full-text data of academic papers to train the word vectors of the methodology entities. We then implemented the entity clustering technique to expand the basic methodology taxonomy to construct a taxonomy with more levels.

This research provides a new concept for constructing a methodology taxonomy of information science. The main differences between the proposed study and previous works can be summarized as follows. In this study, we used a large amount of full-text data from academic papers, and combined manual annotation and entity clustering techniques to build a methodology taxonomy with more levels in a semi-automated manner. This taxonomy can be updated quickly. Whereas traditional three-level methodology taxonomies typically display approximately a dozen abstract research methodologies, our proposed methodology taxonomy shows hundreds of specific research methodology entities that are easier for researchers to understand.

## 2.0 Related Works

Extensive research has been conducted on methodology taxonomies and other aspects related to our work. They can be classified based on whether the focus of research attention is on the construction of methodology taxonomy or the automatic generation of concept hierarchy.

### 2.1 Traditional Methods for Constructing the Methodology Taxonomy of Information Science

The basic concept behind most of the research on the construction of information science methodology taxonomy is as follows. Research methodologies from academic papers related to information science are summarized through content analysis. These research methodologies are then classified based on certain rules to form the final methodology taxonomy. With the increase in related research, the research methodologies summarized by scholars have gradually become consistent. For example, Chu (2015) summarized 16 types of information research methodologies, such as "bibliometric methods," "content analysis methods," and "questionnaire methods". Therefore, the core problem in constructing a methodology taxonomy of information science shifted to classification of the research methodologies. Gradually, some high-impact classification methods were developed; for example, Wang's "hierarchical theory" (Wang 1985) and Miao and Xu's "process theory" '(Miao and Xu 1988).

In Wang's hierarchical theory, methodologies are divided into three major categories: special research methodologies, general research methodologies, and philosophical research methodologies. These three types of research methodologies exhibit parallel relationships at the same level instead of hierarchical relationships. Special research methodologies refers to research methods that are exclusive to information science, including "bibliometrics methods" and "citation analysis methods;" whereas general research methodologies refers to more general research methods that are transplanted and introduced from other disciplines, such as "social survey methods," and "logical reasoning" (Wang 1985).

In the process theory, based on the process for which the method is employed, methodologies are divided into information collection methodologies, information analysis methodologies, and information expression methodologies. Information collection methodologies include "document search," "data collection," "intelligence investigation," "intensive information collection," and "information evaluation." Information analysis methodologies include "fact analysis," "numerical analysis," "index analysis," "future trend analysis," and "image analysis." Information expression methodologies include "text expression" and "image expression" (Miao and Xu 1988).

Most of the existing research has been performed manually. However, in this study, we attempted to implement automation technology.

**2.2  Automatic Generation of Concept Hierarchy**

The methodology taxonomies of information science are generally generated by manual methods; few studies have generated methodology taxonomies automatically. Research methodologies can be considered as conceptual entities (Zhang et al. 2018). Therefore, the methods utilized for automatic generation of concept hierarchies can be used to build a methodology taxonomy of information science. Four methods based on lexical syntax pattern matching, dictionary, association rule, and clustering are used for the automatic generation of concept hierarchies.

A method based on lexical syntax pattern matching was proposed by Hearst (1992), wherein the lexical syntax patterns appearing in the text were used to determine the superordinate and subordinate relationships between words. The inter-word relations were then extracted with high accuracy. The dictionary-based method, first proposed by Amsler (1981) and Calzolari (1984), uses a machine-readable dictionary (MRD) to search for lexical relationships. The relationships between the concepts are then reduced through user feedback and other methods. The accuracy of extracting conceptual hierarchical relationships through this method is generally high; furthermore, the method proposed by Calzolari (1984) achieves an accuracy of more than 90%. However, this method is highly dependent on the accuracy of the MRD. The association rule-based method assumes that if two concepts appear in the same sentence, paragraph, chapter, or document, there must be a close relationship between them, which can be determined by their co-occurrence. It should be noted that the association rule can only determine the existence of a relationship between two conceptual entities; it cannot identify the superordinate or subordinate in two conceptual entities. Sanderson and Croft (1999) proposed a hypothesis based on the association rule: if entity B must appear in the document where entity A appears and entity A does not necessarily appear in the document where entity B appears, then entity A is the lower concept among entities A and B. In the clustering-based method, it is considered that similar conceptual entities appear in similar contexts (Tsui et al. 2010), and features can be extracted from the context of conceptual entities to calculate the similarity between them; similar conceptual entities can be subsequently obtained by the clustering technology. Caraballo (1999) used a bottom-up clustering method to cluster conceptual entities and determine the hierarchical relationship between them. Lam et al. (2007) used a hierarchical clustering method to construct the taxonomy of appliances and Janssens et al. (2019) applied a hybrid clustering method to improve existing journal-based subject-classification schemes.

Methods based on lexical syntax pattern matching and dictionary can generally achieve high accuracy, but they rely heavily on the lexical syntax patterns or dictionary constructed earlier. Moreover, the acquisition of lexical syntax patterns or dictionaries is often labor intensive. The accuracies of the association rule-based and clustering-based methods are generally lower than that of the first two methods; however, these methods do not require lexical syntax patterns or dictionaries in advance and their operations are simple and feasible. In this study, based on the methodology entities extracted from the paper corpus, we used a cluster-based method to build the methodology taxonomy (Zha et al. 2018).

## 3.0 Methodology

### 3.1 Research Framework

This study aims to construct a fine-grained methodology taxonomy of information science in a semi-automatic manner. In comparison to the conventional schemes, the method proposed in this study can update the methodology taxonomy more quickly. We add specific research methodology entities to the methodology taxonomy using a clustering algorithm to enrich it.

The research framework of this study is shown in Figure 1. Researchers from the author's research group manually labeled the method categories used in the papers published in the Journal of the China Society for Scientific and Technical Information from 2009 to 2018 (hereinafter called the full-text content corpus of JCSST) and extracted the methodology entities from each paper (Zhang and Zhang 2020). We also collected full-text data from more than 10,000 academic papers in the field of information science and formed a word vector training corpus for the methodology entities (hereinafter called the training corpus). Then, we used the shallow neural network language model

Word2Vec (Mikolov et al. 2013) to perform distributed representation learning on the methodology entities to obtain their word vectors. Next, the full-text content corpus of JCSST was divided according to the method categories used for annotation; the vectors of the methodology entities appearing in papers under the manually annotated three-level categories were clustered using different clustering algorithms. In this study, we used an internal evaluation method to evaluate three clustering algorithms (affinity propagation clustering, agglomerative hierarchical clustering, and K-means clustering) and selected the optimal clustering result. Finally, we combined the three-level methodology taxonomy constructed by manual annotation with the optimal clustering result, and performed manual fusion and optimization to form a methodology taxonomy of information science.

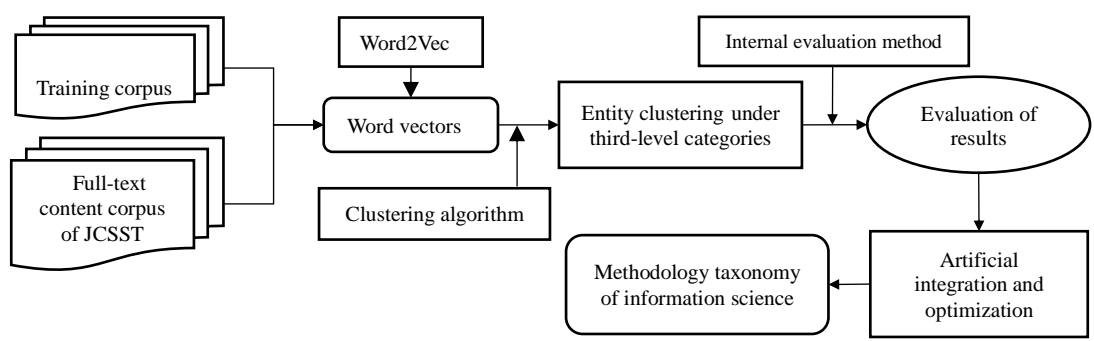

Figure 1. Research framework for the construction of an information science methodology taxonomy.

## 3.2 Experimental Data

Two types of data were used in this study. The first included papers with full-text content from JCSST in the period 2009–2018; we mainly used this data. The second included the full-text content of Chinese academic papers from the field of information science; however, this data was only used to increase the corpus for training the word vectors of methodology entities. The number of papers for these two types of data was 1,349 and 13,720, respectively.

In a series of studies by the author's research group, the researchers manually annotated the research methodology categories of the full-text content corpus of JCSST and constructed a basic three-level methodology taxonomy based on the work of Chu and Ke (2017). In their work, 16 research methodologies were extracted from the research articles published in the Journal of Documentation (JDoc), the Journal of the American Society for Information Science & Technology (JASIS&T), and Library and Information Science Research (LISR). Although the papers published in JCSST were used in our work, we made some adjustments and annotated 21 third-level methodologies. For example, we added "statistical method" and "visualization analysis". If more than one category of methodology was used in a paper, then that paper was labeled under multiple categories; in other words, such papers belonged to multiple research methodology categories. We then counted the number of papers for each methodology category, as listed in Table 1. It is evident that the experimental and statistical methods had the highest number of papers at 739 and 225, respectively.

| First level category | Second level category | Third level category | # papers |
|---|---|---|---|
| Methodology of information science in China | Data collection method | 1: Questionnaire | 71 |
| | | 2: Interview method | 21 |
| | | 3: Delphi method | 3 |
| | | 4: Weblog method | 11 |
| | | 5: Social psychology data collection equipment and methods: eye tracker, etc. | 29 |
| | | 6: Think-aloud | 3 |
| | | 7: Computer aided | 31 |
| | Data analysis method | 8: Bibliometric method | 168 |
| | | 9: Systematic literature review | 0 |
| | | 10: Meta-analysis method | 4 |
| | | 11: Experimental method | 739 |
| | | 12: Comparative research | 25 |
| | | 13: Case analysis | 16 |
| | | 14: Historical analysis | 4 |
| | | 15: Hermeneutics | 166 |
| | | 16: Grounded theory | 8 |
| | | 17: Content analysis | 28 |
| | | 18: Social network analysis | 107 |
| | | 19: Statistical method | 225 |
| | | 20: Visualization analysis | 64 |
| | | 21: Other | 21 |

Table 1. Number of papers in each methodology category.

In addition, the author's research group extracted the methodology entities from each JCSST paper during the period 2009–2018 using a neural network model. To improve the reliability of the extracted results, the authors manually reviewed the methodology entities that appeared more frequently and filtered the results that were misidentified as methodology entities. Approximately 10,000 methodology entities were extracted. The frequency of each methodology entity in 1,349 papers was calculated, and the entities were reviewed in the order of their frequency of appearance. Ultimately, we reviewed the entities with a frequency greater than or equal to 4. The cumulative distribution of the frequency of methodology entities was generated by ranking the reviewed entities and the remaining entities in descending order of their frequency size, as shown in Figure 2.

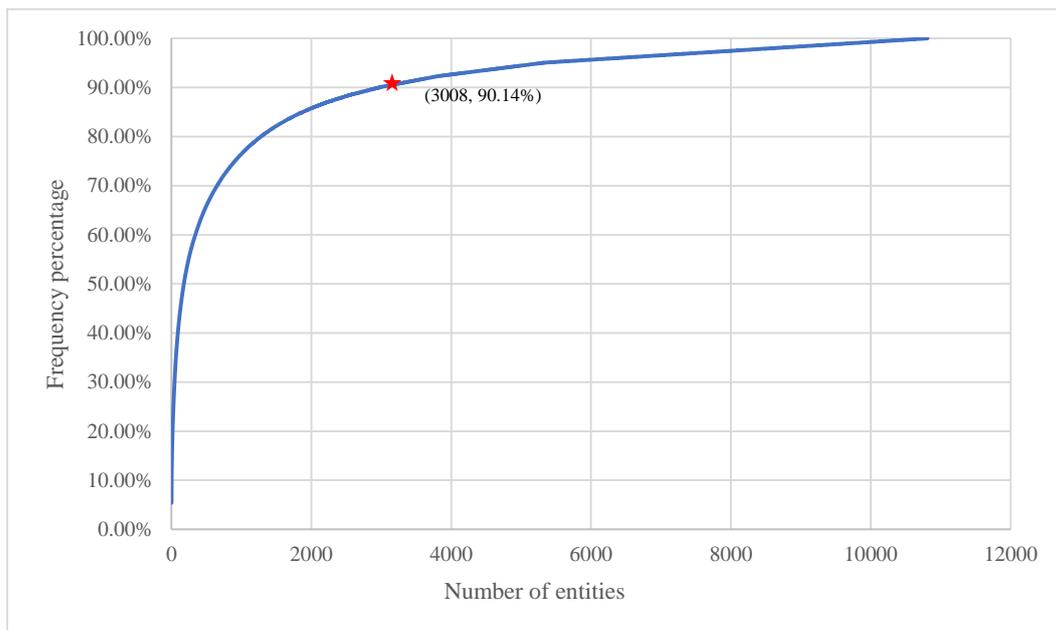

Figure 2. Cumulative distribution of frequencies of methodology entities

In Figure 2, the horizontal axis represents the number of methodology entities ranked in descending order of frequency; the vertical axis represents the cumulative frequency of the methodology entities as a percentage of the total frequency. The top 3008 research methodology entities (i.e., the audited entities with frequencies greater than or equal to 4) constituted 90.14% of the total frequencies, thereby guaranteeing the reliability of more important research methodology entities.

**3.3 Preprocessing of Word Vector Training Corpus**

Before training the word vector of the methodology entities, we must preprocess the training corpus, including sentence and word segmentation.

3.3.1 Sentence Segmentation

Word vector training can be divided into different granularities, such as chapter level, paragraph level, and sentence level. To maximally preserve the semantic relationship between the methodology entities, the word vector was trained in terms of the smallest granularity (sentence level). Therefore, sentence segmentation was required for the training corpus during preprocessing. We used five symbols: "。", "？", "……", "?" and "\n" to segment the sentences.

3.3.2 Word Segmentation

In this research, a word segmentation tool, ICTCLAS (https://github.com/tsroten/pynlpir), was used to segment the training corpus directly. Previous studies also used the ICTCLAS tool to segment words when extracting the methodology entities. ICTCLAS is an excellent Chinese word-segmentation tool with a high accuracy rate.

**3.4 Representation of Informatics Methodology Entities**

We used Word2Vec to train the word vectors of informatics methodology entities. Word2Vec is a shallow neural network language model. The model accepts large amounts of text as input and learns words in a distributed manner to convert them into vectors. These vectors have hundreds of dimensions. When using Word2Vec to train English word vectors, the English text can be directly input as the original data; the model automatically segments the words according to the spaces in the English text. Unlike English text, there are no spaces in Chinese text for word segmentation; therefore, before using Word2Vec for Chinese word vector training, pre-processing is required. In this research, word vector training was performed for the informatics methodology entities at the sentence level. The words obtained after the implementation of word segmentation on each sentence were separated by a space and saved as a line in a text file. We used this file as the input for Word2Vec. The main parameters of Word2Vec include a window size of 5 and 200 vector dimensions.

**3.5 Dividing Informatics Methodology Entities into Corresponding Third-level Categories**

The informatics methodology entities were clustered under the artificially constructed third-level categories. However, they were first divided into corresponding third-level categories. We calculated the chi-square value of the methodology entities for each third-level category and further categorized them according to the largest chi-square value. A total of N categories were determined, of which M belonged to category $C_i$. To examine the correlation between methodology entity t and the third-level category $C_i$, the four observations listed in Table 2 can be used.

| Entity ownership | Belongs to category $C_i$ | Does not belong to category $C_i$ | Total |
|---|---|---|---|
| Contains t | A | B | A + B |
| Does not contain t | C | D | C + D |
| Total | A + C | B + D | N |

Table 2. Four observations between methodology entities and categories.

Then, the chi-square value of methodology entity t and category $C_i$ can be calculated by formula

(1).

$$\chi_{(t,C_i)} = \frac{N(AD-BC)^2}{(A+B)(A+C)(B+D)(C+D)} \quad (1)$$

Finally, the number of methodology entities in each third-level category is listed in Table 3. A total of nine third-level categories with more than 100 methodology entities can be observed. We clustered the methodology entities into these nine categories.

| Third-level category | Number of entities | Third-level category | Number of entities |
| --- | --- | --- | --- |
| 1: Questionnaire | 138 | 12: Comparative research | 125 |
| 2: Interview method | 53 | 13: Case analysis | 44 |
| 3: Delphi method | 42 | 14: Historical analysis | 7 |
| 4: Weblog method | 46 | 15: Hermeneutics | 200 |
| 5: Social psychology data collection equipment and methods: eye tracker, etc. | 62 | 16: Grounded theory | 37 |
| 6: Think-aloud | 13 | 17: Content analysis | 63 |
| 7: Computer aided | 115 | 18: Social network analysis | 218 |
| 8: Bibliometric method | 167 | 19: Statistical method | 111 |
| 10: Meta-analysis method | 51 | 20: Visualization analysis | 153 |
| 11: Experimental method | 1252 | 21: Other | 65 |

Table 3. Number of methodology entities in each third-level category.

**3.6 Using the Clustering Method to Construct Deep Methodology Taxonomy**

Based on the three-level methodology taxonomy constructed by manual annotation, we used AP (affinity propagation) clustering (Frey and Dueck 2007), agglomerative hierarchical clustering (Brandes et al. 2003; Meter 2019), and K-means clustering (Hartigan and Wong 1979) to cluster the methodology entities. Finally, the optimal clustering result was selected to deepen the hierarchy of the methodology taxonomy.

3.6.1. Using the AP clustering algorithm to cluster methodology entities

In the AP clustering algorithm, the similarity matrix of all entities was considered as input. We represented methodology entities as low-dimensional vectors by Word2Vec. In order to measure the distance of methodology entities in vector space, cosine similarity was used. We calculated the cosine similarity between methodology entities to construct a similarity matrix. During the AP clustering process, two types of information are transmitted: attracting information and belonging information. The similarity between two entities can be regarded as the degree of attraction or attribution. For example if Entity A has a strong appeal to entity B, it implies that both are similar. Alternatively, entity B has a strong sense of belonging to entity A. Preference is a very important parameter in AP clustering. It denotes the diagonal values of the similarity matrix, also known as the reference degree and describes the probability that each point becomes the cluster center. The value of preference determines the final number of clusters; generally, the smaller the value of preference, the smaller the number of clusters.

In this study, we set multiple groups of preference values to cluster the methodology entities and evaluate the clustering results of each group to select the optimal clustering result. For each cluster of methodology entities in the clustering results, we selected a representative methodology entity as the tag of the cluster or manually assigned a tag for the cluster. The cluster tag and methodology entities in the cluster form a hierarchical relationship, thereby achieving the purpose of deepening the hierarchy of methodology taxonomy.

3.6.2. Using an agglomerative hierarchical clustering algorithm to cluster methodology entities

The agglomerative hierarchical clustering algorithm is a bottom-up clustering approach. In this method, first, each entity is treated as a cluster and a similarity measure (such as cosine similarity) is selected. Then, two entities with the highest similarity are selected and merged into a cluster at each iteration. Further similarity comparisons are made until all entities are merged into one cluster

(Brandes et al. 2003). In this study, the agglomerative hierarchical clustering algorithm was also used after obtaining the vectors of the methodology entities. First, we considered each methodology entity as a cluster and calculated their cosine similarities. Clusters were merged and the center vector of all methodology entities in each cluster was used as the representative vector of the cluster until all clusters were merged into one cluster. This clustering resulted in a tree structure. We divided the tree clustering results at different hierarchical nodes to obtain clustering results with different numbers of clusters. We then evaluated them to obtain the best results.

3.6.3 Using K-means clustering algorithm to cluster methodology entities

The following steps were used in the K-means clustering algorithm for clustering informatics methodology entities.

Step 1: Select k entities as the initial clustering centers from all methodology entities.

Step 2: Calculate the distance between the remaining methodology entities and the k cluster centers and merge the remaining methodology entities into their closest cluster centers.

Step 3: Calculate the average vector of all methodology entities in each cluster as the new cluster center.

Step 4: Repeat Steps 2 and 3 until the methodology entities in all clusters do not change, or manually set the number of clustering rounds. After n rounds of clustering, clustering is stopped. The obtained result is given as the output.

The value of k has a significant impact on the clustering results. Zhu et al. (2009) first used AP clustering to obtain a clustering result. In their work, the number of clusters in the AP clustering result was used as the k value to optimize K-means clustering. In this study, we set the K-means clustering parameter, k, with reference to the number of clusters of the optimal clustering results in

AP clustering and agglomerative hierarchical clustering. We also trialed multiple sets of different k values to cluster the methodology entities and evaluated their results to select the optimal clustering result.

## 4.0 Results

**4.1 Evaluation Indicator of Clustering Results**

In this study, we attempted to use three clustering methods to cluster informatics methodology entities. To select the best clustering method, we used the Silhouette coefficient (Aranganayagi and Thangavel 2007) to evaluate the clustering results.

The formula for calculating the Silhouette coefficient of entity i is as follows:

$$S_i = \frac{b_i - a_i}{\max\{a_i, b_i\}}, \qquad (2)$$

where $S_i$ represents the contour coefficient of entity i, $a_i$ represents the average value of the degree of dissimilarity between entity i and other entities in the same cluster, $b_i$ represents the minimum value of the degree of dissimilarity between entity i and other clusters and $\max\{a_i, b_i\}$ represents the larger of the two values. Generally, the average value of the Silhouette coefficient of all entities is used to evaluate the clustering results. The larger the average Silhouette coefficient, the better the clustering effect.

**4.2 Evaluation and Comparison of the Three Clustering Algorithms**

The methodology entities in the papers of each third-level category were clustered. To ensure the clustering effect, we selected nine third-level categories with more than 100 methodology entities for clustering. We used AP clustering, hierarchical clustering, and K-means clustering. Additionally, we used the Silhouette coefficient as an indicator to evaluate and compare the clustering results of different algorithms to determine the optimal one.

4.2.1 Evaluation of AP Clustering

While performing AP clustering on the informatics methodology entities in each category, the preference value was set to vary from 0 to 1 at 0.05 intervals. In this study, we used the Silhouette coefficient to measure the clustering effect under different preference values. However, in the calculation of the Silhouette coefficient, we must ensure that the number of clusters be less than the number of entities. Simultaneously, to ensure a sufficient number of clusters, we only recorded the preference values and the corresponding Silhouette coefficients for clusters greater than or equal to 10 and less than the number of entities.

Taking the experimental method as an example, the Silhouette coefficients under different preference values when performing AP clustering on the methodology entities are shown in Figure 3. It can be observed that when the value of the preference parameter is set to 0.5, the Silhouette coefficient is the largest and the clustering effect is best.

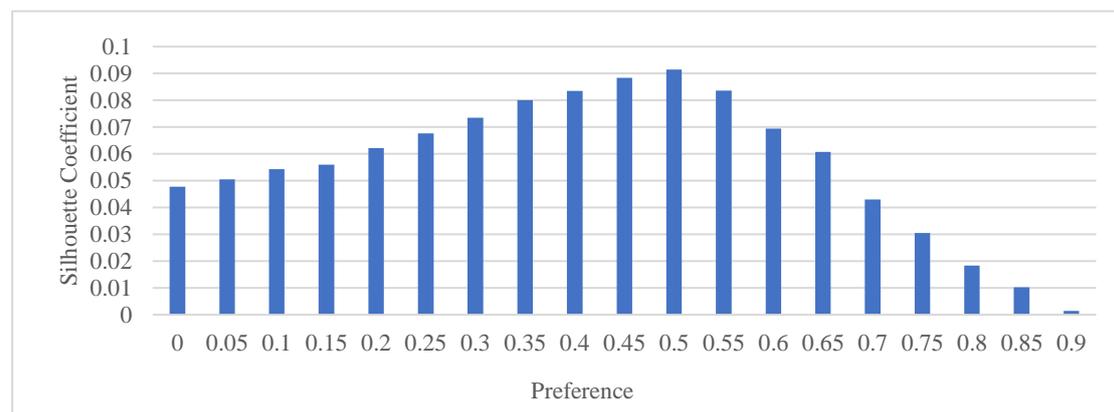

Figure 3. Silhouette coefficients of AP clustering results under different preference values (experimental method).

4.2.2 Evaluation of Agglomerative Hierarchical Clustering

When using an agglomerative hierarchical clustering algorithm to cluster the informatics methodology entities in each category, we first obtained a clustering tree and divided it into clusters according to different similarity levels. In this study, we divided the clusters with similarity values ranging from 0 to 1 at 0.02 intervals. Similarly, to ensure a sufficient number of clusters, we only recorded the results in which the number of clusters was greater than or equal to 10. Then, we calculated the Silhouette coefficients for each cluster result.

Taking the experimental method as an example of a methodology entity, agglomerative hierarchical clustering was used to cluster the methodology entities. The Silhouette coefficients of the clustering results divided by different similarity levels are shown in Figure 4. It is evident that when the clusters were classified at a similarity level of 0.5, the Silhouette coefficient was the largest and the best clustering effect was observed.

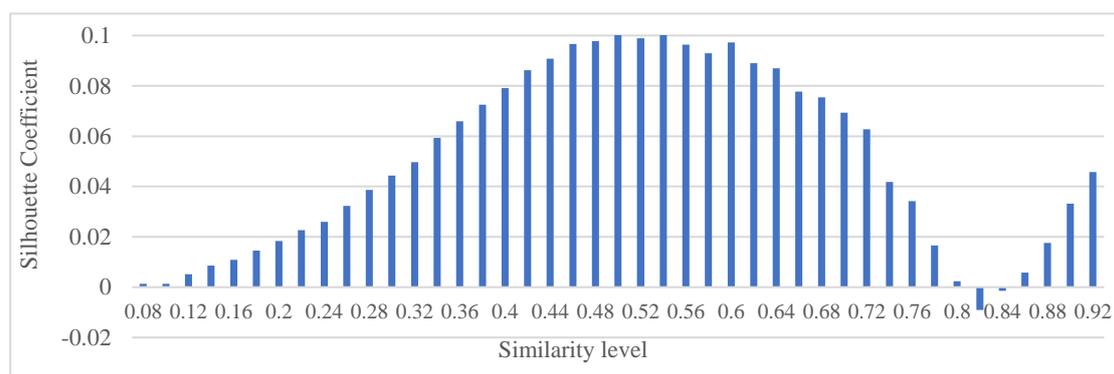

Figure 4. Silhouette coefficients of agglomerative hierarchical clustering results divided by different similarity levels (experimental method)

4.2.3 Evaluation of K-means Clustering

When the K-means clustering algorithm was used to cluster the informatics methodology entities in each category, the clustering results were mainly affected by the number of initial clustering centers, K. By observing the optimal clustering results of AP clustering and agglomerative hierarchical clustering, we found that the number of clusters varied between 10 and 120. Therefore, the values of K were selected to be from 10 to 120 in intervals of 10. Subsequently, K-means clustering was performed 13 times and the Silhouette coefficient of each clustering result was calculated.

Taking the experimental method as an example, K-means clustering was performed on the methodology entities. The Silhouette coefficients of the clustering results at different K values are shown in Figure 5. It is evident that when the value of K is 10, the Silhouette coefficient is the largest and the best clustering effect is observed.

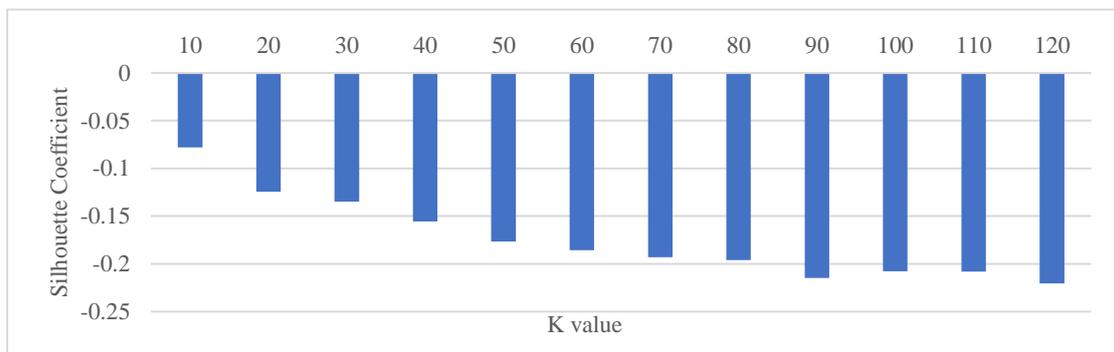

Figure 5. Silhouette coefficients of K-means clustering results at different K values (experimental method)

4.2.4 Selection of the Optimal Clustering Algorithm

Finally, we must select the best clustering method among the three to build our taxonomy. Accordingly, we compared the Silhouette coefficients of the three clustering methods in different situations, shown in Figure 6. The vertical axis represents the Silhouette coefficients of the three clustering methods that achieved the optimal clustering effect in different categories. As evident from Figure 6, the Silhouette coefficients of AP clustering and agglomerative hierarchical clustering are significantly larger than those of K-means clustering in different situations. K-means clustering has the worst effect; whereas the effect of agglomerative hierarchical clustering is slightly better than that of AP clustering. Therefore, we used the results of agglomerative hierarchical clustering to build a methodology taxonomy of information science.

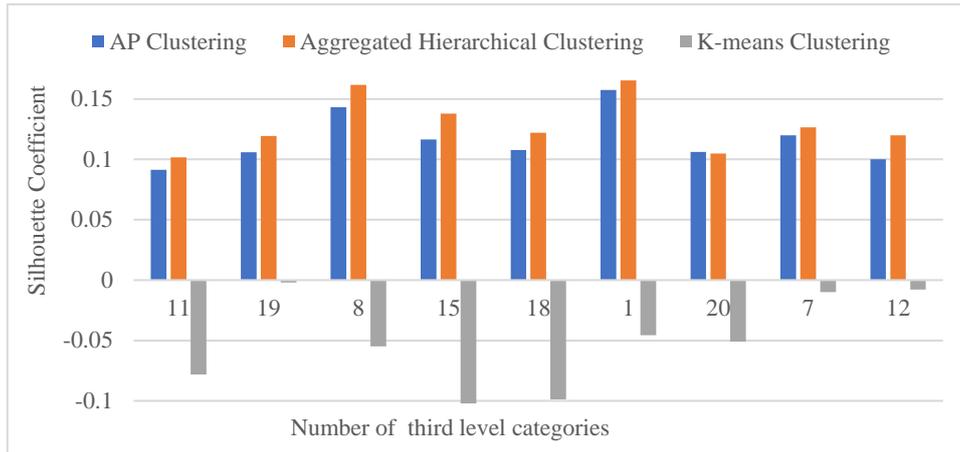

Figure 6. Silhouette coefficients for the best results of the three clustering methods

**4.3 Optimal Result Analysis of Agglomerative Hierarchical Clustering**

We used the experimental method and social network analysis with the largest number of methodology entities as examples to analyze the results of agglomerative hierarchical clustering.

We evaluated the sum of the frequency of all methodology entities in each cluster in their corresponding category as the impact of that cluster, ranked the clusters by the impact size, and analyzed the top five clusters with the greatest impact. Each cluster was labeled according to the frequency of the methodology entities or the commonality of most methodology entities. The top five clusters with the greatest impact on the optimal clustering results in the experimental method are summarized in Table 4. The methodology entities in each cluster are sorted according to their frequency.

| No. | Sum of frequency of entities | Cluster tag | Methodology entities |
|---|---|---|---|

| 1 | 4820 | Experiment | experiment, comparative experiment, experimental verification, simulation experiment, experimental comparison |
| 2 | 1790 | Evaluation indicators | recall rate, f value, recall rate, accuracy rate, classification accuracy rate, average accuracy rate, precision rate, average precision rate, f1 value, f-measure value, average recall rate |
| 3 | 1084 | Classification model | svm, support vector machine, classification model, naive Bayes, svm classifier, svm model, svm method, svm algorithm, support vector machine, svm classification, support vector machine model |
| 4 | 863 | Statistical learning model | crf, conditional random field model, crfs, conditional random field, hmm, maximum entropy model, hidden Markov model, machine learning model, crf model, annotation model, crfs model, statistical learning method, conditional random fields, memm, statistical learning model |
| 5 | 833 | Domain ontology | domain ontology, ontology method, ontology, semantic model |

Table 4. Top five clusters with the greatest impact in the experimental method.

Table 4 demonstrates that some of the top five clusters with the greatest impact in the experimental method are clustered in different forms of the same research method entity. Additionally, some of the methodology entities with certain commonalities are also clustered together. The table also reflects the reliability of the clustering effect to a certain extent. An example of a cluster wherein methodology entities are clustered in different forms of the same research method entity is represented by cluster 1 with the highest impact, whose cluster label is "Experiment." "Experiment" is consistent with the category of experimental method. The methodology entities in this cluster are mostly related to "experiment," such as "experiment," "comparative experiment," and "simulation experiment. An example of a cluster wherein methodology entities with certain commonalities are clustered together is represented by clusters 2 and 3, which contain different types of methodology entities. Some indicators of cluster 2 can be used to evaluate the effect of the experiment, such as the "recall rate," "f value," and "correct rate." The methodology entities in cluster 3 belong to classification models, such as "svm," "naive Bayes,"

and "k-nn."

| No. | Sum of frequency of entities | Cluster tag | Methodology entities |
|---|---|---|---|
| 1 | 706 | Centrality index | centrality, centrality, intermediate centrality, intermediate centrality, near centrality, intermediate centrality, intermediate centrality, point centrality, degree centrality, degree centrality, degree centrality, feature vector centrality, betweenness centrality, intermediary centrality, eigenvector centrality, centrality index, intermediate centrality index, proximity centrality |
| 2 | 328 | Social network analysis | social network analysis, social network analysis method, graph theory, network analysis, complex network theory, complex network analysis method, network analysis method, citation network analysis, social network analysis theory, complex network method |
| 3 | 300 | Visualization tools | ucinet, network diagram, network visualization, ucinet software, pajek, netdraw software, pajek software, gephi software, social network analysis software, sna analysis, network analysis tools |
| 4 | 207 | Network indicators | network density, average path length, average clustering factor, cohesion index, average shortest path length |
| 5 | 182 | Coupling analysis | coupling, literature coupling |

Table 5. Top 5 clusters with the greatest impact in social network analysis.

The top five clusters with the greatest impact on the optimal clustering results of social network analysis are listed in Table 5. Similar to the optimal clustering results of the experimental method, the five clusters are either different variants of one methodology entity or multiple methodology entities with common characteristics. The cluster "social network analysis," which corresponds to the social network analysis method, has the second highest impact. Cluster 1, which has the highest impact, contains some variants of "centrality," but "centrality" is often measured during network analysis. Therefore, this situation is also reasonable. In addition, cluster 3 contains multiple types of methodology entities, but their commonality is that they can be used for network analysis and visualization.

**4.4 Construction Results for the Methodology Taxonomy of Information Science**

Evaluating and comparing the effects of the three clustering algorithms, the effect of agglomerative hierarchical clustering was found to be the best. Therefore, based on a three-level methodology taxonomy formed by manual annotation, we used the results of agglomerative hierarchical clustering as a supplement. We selected the top five clusters with the highest impact in agglomerative hierarchical clustering for the nine categories and artificially assigned a tag to each cluster as the fourth hierarchical level. Subsequently, the collection of methodology entities in each cluster was considered the fifth level. Thus, we developed a methodology taxonomy of information science with five levels. The top four levels of taxonomy are shown in Figure 7. The entire taxonomy can be viewed at https://chengzhizhang.github.io/research/methodology_taxonomy/mtis.html.

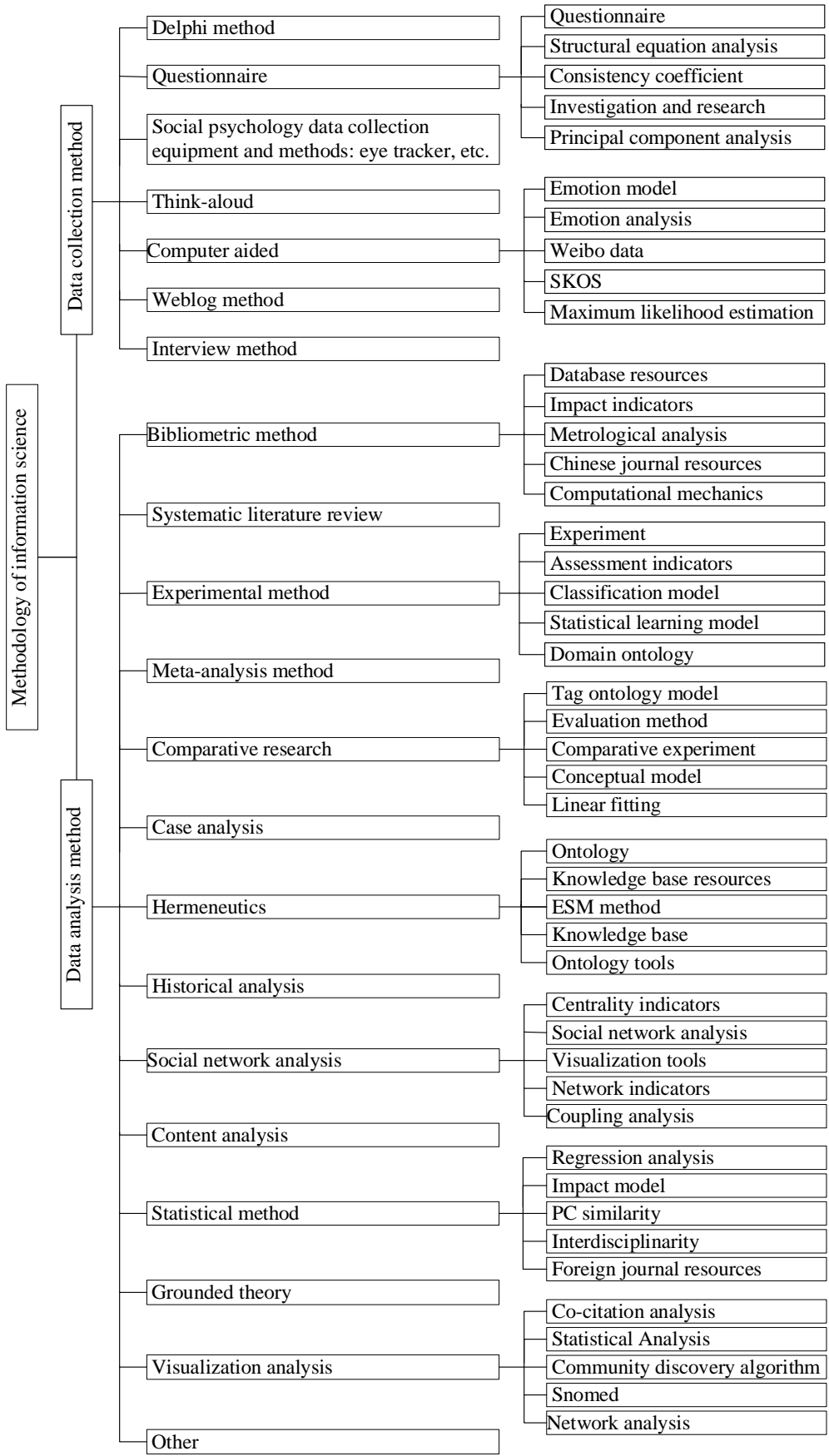

Figure 7. Methodology taxonomy of information science in China

As can be seen in Figure 7, in comparison to the traditional three-level methodology taxonomy, our five-level methodology taxonomy divides informatics methodologies more finely. We subdivided the five fourth-level categories under each third-level category. The methodology entities included in each fourth-level category can also be observed. In terms of the number of research methodologies, the traditional three-level methodology taxonomy contains only a dozen abstract research methodologies. However, the methodology taxonomy constructed in this study contains hundreds of methodology entities. Moreover, the types of methodology entities are diverse, including algorithms, models, databases, indicators, systems, and tools, which are detailed and specific.

Observing the five-level methodology taxonomy constructed in this study, we found that the fifth level of the methodology taxonomy contains several different forms of methodology entities, which can help researchers in obtaining more comprehensive and relevant literature by searching different keywords. The use of the chi-square method in this study for dividing the methodology entities into different third-level categories implies that the divided methodology entities are more reflective of the characteristics of the categories to which they belong. In fact, many methodology entities are used in multiple third-level categories. For example, the methodology entities related to "experiment" and "statistics" appear in almost all third-level categories. They are called "popular research methods" or "universal research methods" in information science research.

## 5.0 Summary and Future Works

In this study, the full-text content of 1,349 papers published in JCSST from 2009 to 2018 was used to build a methodology taxonomy of information science. In previous works, the methodology categories were annotated to extract the methodology entities, and a basic three-level methodology taxonomy was constructed manually. In this study, we first trained the word vectors of the methodology entities and selected nine third-level categories comprising the largest number of methodology entities after feature selection. We used three clustering methods to cluster the methodology entities with high frequency in each category and selected the best clustering method by calculating the Silhouette coefficient. Then, we expanded the basic three-level methodology taxonomy with the results of agglomerative hierarchical clustering and constructed a five-level methodology taxonomy of information science. The traditional three-level methodology taxonomy is relatively abstract. The five-level methodology taxonomy constructed in this study displays more details of the informatics methodologies, which contain hundreds of specific methodology entities. Thus, this taxonomy can be easily understood by researchers.

In comparison to the existing research on the construction of methodology taxonomy in information science, the innovations of this research are as follows. Whereas traditional methods use a limited number of academic papers or only specific parts of the papers, this study used a significantly larger number of academic papers. The relevant sections of the academic papers contain numerous descriptions of the research methodologies, and this study utilized this information. Most traditional studies implement manual labeling, which is time-consuming and labor-intensive; moreover, the methodology taxonomy thus obtained cannot be updated quickly. This research constructed an information science methodology taxonomy in a semi-automated manner that can be updated quickly.

However, this study has certain limitations. The process of manual labeling is subjective and depends on the background of professional knowledge; thus, it cannot be absolutely accurate. This research directly used the results of methodology entity extraction in the related work of the author's research group. Although the author manually proofread the frequent methodology entities to ensure accuracy, not all research method entities could be extracted. This may have led to the omission of certain methodology entities and affected the clustering results.

In future work, we plan to optimize the manual labeling process to make it more efficient, and improve the accuracy of methodology entity extraction and the clustering effect. In addition, we only collected 1,349 papers published in JCSST from 2009 to 2018 to construct the methodology taxonomy. Although JCSST is a representative journal of information science in China, the methodology taxonomy we constructed may still deviate slightly from reality. Thus, we could expand the variety of information science journals and increase the number of papers to make the results closer to reality.

## Acknowledgements

This work is supported by National Natural Science Foundation of China (Grant No. 72074113, Grant No.71974095), Major Projects of National Social Science Fund (Grant No. 16ZDA224).